\documentclass[journal=nalefd,manuscript=letter]{achemso}

\usepackage{graphicx}
\usepackage{epstopdf}
\usepackage{braket}
\usepackage{chemformula} % Formula subscripts using \ch{}
\usepackage[T1]{fontenc} % Use modern font encodings
\usepackage[version=3]{mhchem} % Formula subscripts using \ce{}
\usepackage{etoolbox}% needed for the patch  <<<

\makeatletter
\renewcommand*{\acs@author@fnsymbol@symbol}[1]{% Use numbers instead of symbols, * is for email
    \ifcase #1 *\or
    1\or
    2\or
    3\or
    4\or
    5\or
    6\or
    7\or
    8\or
    \dagger\or
    \ddagger\or
    \fi}

\usepackage{atbegshi}% http://ctan.org/pkg/atbegshi
\AtBeginDocument{\AtBeginShipoutNext{\AtBeginShipoutDiscard\addtocounter{page}{-1}}}

\author{Jun Gao}
\email{junga@kth.se}
\affiliation{KTH Royal Institute of Technology, Department of Applied Physics, Albanova University Centre, Roslagstullbacken 21, 106 91 Stockholm Sweden.}

\altaffiliation{Corresponding Author}
\altaffiliation{Equal contribution}

\author{Govind Krishna}
\email{govindk@kth.se}
\affiliation{KTH Royal Institute of Technology, Department of Applied Physics, Albanova University Centre, Roslagstullbacken 21, 106 91 Stockholm Sweden.}

\altaffiliation{Corresponding Author}
\altaffiliation{Equal contribution}

\author{Edith Yeung}
\affiliation{National Research Council of Canada, Ottawa, Ontario, Canada, K1A 0R6.}
\alsoaffiliation{University of Ottawa, Ottawa, Ontario, Canada, K1N 6N5.}
\altaffiliation{Equal contribution}

\author{Lingxi Yu}
\affiliation{National Research Council of Canada, Ottawa, Ontario, Canada, K1A 0R6.}
\altaffiliation{Equal contribution}

\author{Sayan Gangopadhyay}
\affiliation{Institute for Quantum Computing and Department of Physics and Astronomy, University of Waterloo, Waterloo, Ontario N2L 3G1, Canada}
\altaffiliation{Equal contribution}

\author{Kai-Sum Chan}
\affiliation{National Research Council of Canada, Ottawa, Ontario, Canada, K1A 0R6.}
\alsoaffiliation{Electrical and Computer Engineering, University of Toronto, Toronto, Ontario M5S 3G4, Canada}
\alsoaffiliation{Quantum Bridge Technologies Inc., 108 College Street, Toronto, Canada M5G 0C6}

\author{Chiao-Tzu Huang}
\affiliation{Department of Electrophysics, National Yang Ming Chiao Tung University, Hsinchu 30010, Taiwan}

\author{Thomas Descamps}
\affiliation{KTH Royal Institute of Technology, Department of Applied Physics, Albanova University Centre, Roslagstullbacken 21, 106 91 Stockholm Sweden.}

\author{Michael E. Reimer}
\affiliation{Institute for Quantum Computing and Department of Physics and Astronomy, University of Waterloo, Waterloo, Ontario N2L 3G1, Canada}
\alsoaffiliation{Institute for Quantum Computing and Department of Electrical and Computer Engineering, University of Waterloo, Waterloo, Ontario N2L 3G1, Canada}

\author{Philip J. Poole}
\affiliation{National Research Council of Canada, Ottawa, Ontario, Canada, K1A 0R6.}

\author{Dan Dalacu}
\affiliation{National Research Council of Canada, Ottawa, Ontario, Canada, K1A 0R6.}
\alsoaffiliation{University of Ottawa, Ottawa, Ontario, Canada, K1N 6N5.}
\alsoaffiliation{Electrical and Computer Engineering, University of Toronto, Toronto, Ontario M5S 3G4, Canada}

\author{Val Zwiller}
\affiliation{KTH Royal Institute of Technology, Department of Applied Physics, Albanova University Centre, Roslagstullbacken 21, 106 91 Stockholm Sweden.}

\author{Ali W. Elshaari}
\affiliation{KTH Royal Institute of Technology, Department of Applied Physics, Albanova University Centre, Roslagstullbacken 21, 106 91 Stockholm Sweden.}
\email{elshaari@kth.se}
\altaffiliation{Corresponding Author}

\title{On demand single photon generation and coherent control of excitons from resonantly driven nanowire quantum dots}

\abbreviations{IR,NMR,UV}

\begin{document}

% \begin{tocentry}

% Some journals require a graphical entry for the Table of Contents.
% This should be laid out ``print ready'' so that the sizing of the

% Inside the \texttt{tocentry} environment, the font used is Helvetica
% 8\,pt, as required by \emph{Journal of the American Chemical

% The surrounding frame is 9\,cm by 3.5\,cm, which is the maximum
% permitted for  \emph{Journal of the American Chemical Society}
% graphical table of content entries. The box will not resize if the
% content is too big: instead it will overflow the edge of the box.

% This box and the associated title will always be printed on a
% separate page at the end of the document.

% \end{tocentry}

\begin{abstract}

Coherent control of single photon sources is a key requirement for the advancement of photonic quantum technologies. Among them, nanowire-based quantum dot sources are popular due to their potential for on-chip hybrid integration. Here we demonstrate on-demand single-photon generation ($g^{(2)}(0)(X^{*}) =0.078$ and $g^{(2)}(0)(X)= 0.03$) from resonantly excited InAsP/InP nanowire quantum dots and observe Rabi oscillations in the dot emission, indicating successful coherent manipulation of the excitonic states in the nanowire. We also measure a low emission time jitter for resonant excitation as compared to above-band excitation. This work addresses the long-standing challenge of resonantly exciting nanowire-quantum dots. It paves the way for hybrid quantum photonic integration, enabling spin-photon entanglement and matter memories on-chip.

%This is an abstract for the \textsf{achemso} document class demonstration document.  An abstract is only allowed for certain manuscript types.  The selection of \texttt{journal} and \texttt{manuscript} will determine if an abstract is valid.  Ifnot, the class will issue an appropriate error.
\end{abstract}


Single photons are deemed an excellent choice of quantum bits (qubits) due to their distinct advantages over other candidates, such as low decoherence, the possibility of quantum state manipulation at room temperature, and ease of integration with the existing global optical fiber network. Advancements in photonic quantum technologies demand efficient methods to generate time-tagged, coherent, indistinguishable single photons at high repetition rates with directional emission and precise nanoscale positioning capabilities. Quantum dots (QDs) embedded in nanowires have emerged as promising candidates satisfying these demands, along with their potential for integration into scalable quantum photonic systems\cite{Claudon2010, Dalacu2012, Versteegh2014, elshaari2020hybrid, Zadeh2016, Descamps2023,gao2023scalable,chang2023nanowire}. Such nanowire-based quantum emitters also offer additional advantages like the possibility of multiple QDs integration into a single nanowire (enabling advanced multi-qubit devices)\cite{multipleQDsingleNW}, precise control of size and position of the emitters (also enabling wavelength tuning)\cite{multipleQDsingleNW, Patrik2022, Zadeh2016, Subannajui2011}, high extraction efficiency\cite{Claudon2010, Reimer2012} and maximal coupling to optical fibre modes\cite{Munsch2013, Bulgarini2014}. While significant progress has been made in other QD systems~\cite{ding2023high,tomm2021bright,somaschi2016near,he2013demand}, sub-Poissonian statistics under resonant excitation of nanowire quantum dots have not yet been demonstrated~\cite{leandro2020resonant, Lagoudakis2016}. Our work addresses these gaps, as coherently driven nanowire quantum dot sources can enable spin-photon entanglement and the development of optical memories, representing a significant leap forward in the field of hybrid quantum photonic integration.  

The usual incoherent methods of pumping a QD, such as above-band excitation, lead to high levels of dephasing due to uncontrollable emission time jitter for relaxation from the upper levels to the s-shell\cite{Troiani2006, Kaer2013}, homogeneous broadening of the emission due to carrier-carrier and carrier-phonon interactions within the exciton lifetime, and inhomogeneous broadening due to charge noise from the QD surroundings\cite{Bennett2005, Kuhlmann2013, Reimer2016}. Resonant s-shell excitation is a remedy for such dephasing effects\cite{Kiraz2004, he2013demand}. Resonant fluorescence (RF) by pulsed laser excitation has a distinct advantage over continuous-wave laser excitation in generating precisely time-tagged and synchronized single photons. Such synchronized single photon emitting sources find direct application in many quantum information protocols\cite{Pan2012, Jeremy2009, Kok2007}. 

In this work, we perform resonant s-shell excitation of the neutral and charged exciton emission lines of an InAsP/InP nanowire quantum dot. Above-band excitation at powers much below the saturation level reveals three distinct emission lines. To identify each line, we perform cross-correlation measurements between pairs of excitonic complexes using a transmission spectrometer based setup. Second-order autocorrelation measurements on RF excitation yielded a $g^{(2)}(0)$ value of 0.078 for the charged exciton and 0.21 for the neutral exciton. We observe Rabi oscillations of the charged exciton complex with varying pump power, indicating coherent control of the emission. With another similar sample with a lower exchange splitting (on a comparable setup), we obtained a $g^{(2)}(0) = 0.03$ and observed Rabi oscillations for the neutral exciton. We observed a lower emission time jitter for resonant excitation than above-band excitation. We also compare the  $g^{(2)}(0)$ characteristics of the resonant and above-band excitation plots to discuss the importance of pump laser rejection in a resonant excitation scheme.

\begin{figure}[t!]
\centering\includegraphics[width=\linewidth]{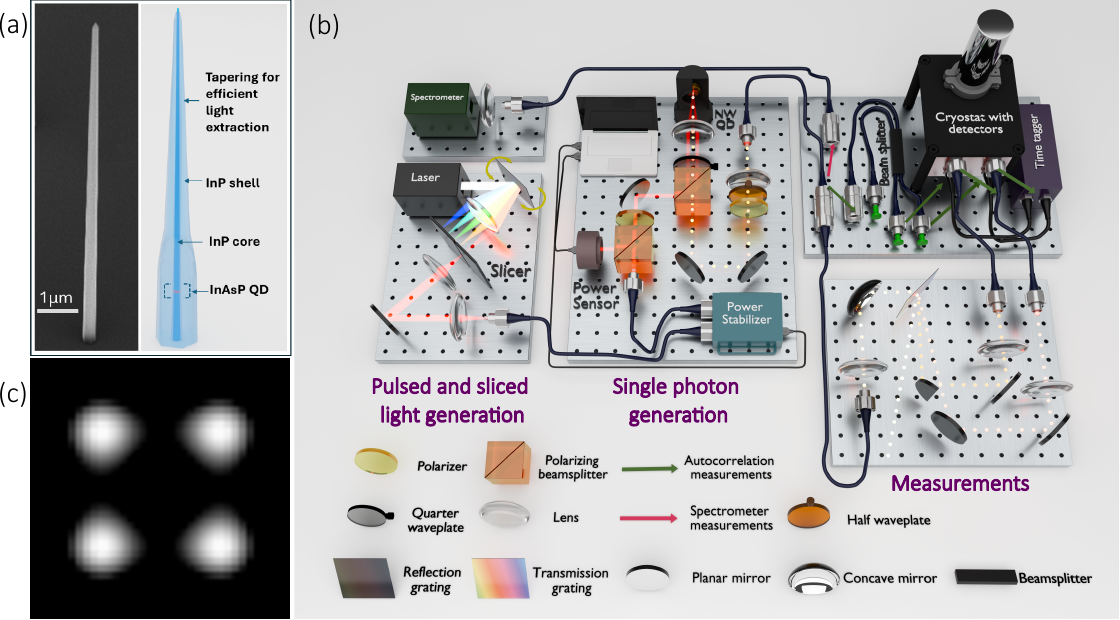}
\caption{(a) SEM of the nanowire (left) and a schematic of the nanowire structure (right). The QD is not visible in the SEM image as it is embedded within the nanowire. (b)  Schematic of the experimental setup. The measurements sections of the image (The two sections at the rightmost, also labelled "Measurements" in the figure) shows the scheme for cross-correlation measurements. The green and pink arrows denote the fiber re-connections for doing auto-correlation and spectrometer measurements, respectively. For the above-band excitation, the slicer is bypassed, and the light from the laser is directly sent to the cryostat with the nanowire. (c) Numerically simulated "flower" like profile of the pump laser reflected from the nanowire under resonant excitation. }
\label{fig1}
\end{figure}

Each quantum emitter sample studied in this manuscript consist of an InAsP QD embedded in an InP nanowire ((Fig.~\ref{fig1}(a)), which is placed in a cryostat operating at 4\,K. The disc-like QD structure embedded within the nanowire has a thickness of \(\sim \) 5nm and a diameter of \(\sim \) 20 nm. Fig.~\ref{fig1}(b) shows the experimental setup for the excitation of the nanowire QD and the measurement of the dot emission. The nanowire is excited using a 80 MHz (except for the measurement in Fig.~\ref{NRC_plots}(a) discussed later) pulsed laser, also equipped with a reflection grating based slicer setup that enables picosecond range tunability of the pulse length. The pump laser power is monitored and stabilized by a power sensor and a MEMS-based power stabilizer, respectively. The light is then linearly polarized and reflected from a polarizing beam splitter (PBS) before being shined onto the nanowire in the cryostat.

The pump rejection is achieved along the emission path through a combination of polarization and spatial filtering. The polarization filtering is accomplished by creating cross-polarization between the emitted single photons and the pump laser, with the help of a zero-order quarter wave plate and subsequent passage through a polarizing beam splitter and a linear polarizer. In contrast to self-assembled QDs, in the case of nanowire QDs, the pump filtering based on a cross-polarization scheme does not suffice as the irregular surface and multiple facets on the nanowire distort the scattered light and make its polarization non-uniform~\cite{leandro2020resonant}. So a small fraction of the pump light will remain after polarization filtering, which will be rejected by spatial filtering. The spatial filtering is achieved by coupling the emission from the cryostat to a single-mode fiber, which rejects the flower-shaped profile of the reflected laser beam ~\cite{leandro2020resonant} and accepts the Gaussian QD emission profile. With the flower-like profile, the pump light intensity at the center of the optical path cross-section (where the intensity of the Gaussian QD emission is highest) is reduced to almost zero. The single-mode fiber then acts as a pinhole, which collects only the dot emission and rejects the pump. This setup effectively removes the scattered laser light, achieving a laser suppression exceeding 80\,dB. This pump rejection step is vital and is one of the major limiting factors for RF experiments, as in these experiments, the pump light wavelength is the same as the dot emission wavelength, eliminating the possibility of using bandpass filters. The stability of the dark-field microscope is maintained over extended periods, allowing for continuous operation without the need for realignment~\cite{kuhlmann2013dark}. The QD emission is then sent to either a spectrometer for photoluminescence measurements or to superconducting nanowire single-photon detectors (SNSPDs) for autocorrelation or cross-correlation measurements.

\begin{figure*}[t!]\centering\includegraphics[width=1\linewidth]{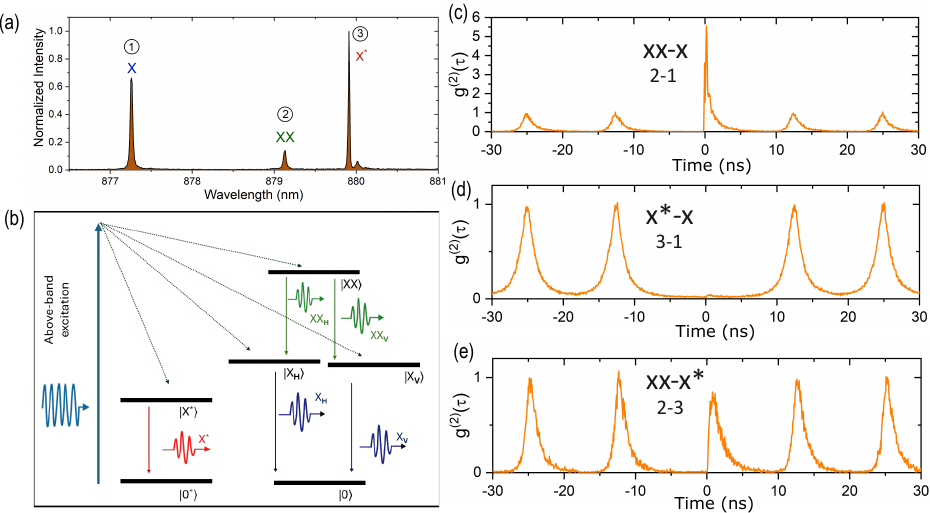}
\caption{(a) Photoluminescence (PL) spectrum of the nanowire QD at above-band non-resonant excitation (using 790 nm, 80 MHz pulsed laser). (b) Schematic of the energy levels and above-band excitation of a QD. (c-e) Cross-correlation measurements between pairs of excitonic complexes using above-band excitation at power $P_{sat}/5$ ($P_{sat}$ = 600 nW).}
\label{fig2}
\end{figure*}

Figure~\ref{fig2}(a) shows the measured photoluminescence (PL) spectrum at above-band excitation (790\,nm) of the nanowire QD. The nanowire QD at low excitation power shows three distinct emission lines: a neutral exciton ($X$), a charged exciton ($X\textsuperscript{*}$), and a neutral biexciton ($XX$). These emissions are generated as per the scheme shown in Figure~\ref{fig2}(b). The neutral exciton $X$ exhibits fine structure splitting due to anisotropic exchange interaction in the dot. The fine structure splitting of the exciton is not visible in the PL spectrum due to the limited resolution of our spectrometer.

%This splitting is vital for generating polarization-entangled photon pairs ($XX_V, X_V$ and $XX_H, X_H$).

%To verify the nature of each emission line, we perform polarization state analysis of the emitted photons, excitation power dependence measurements, and cross-correlation measurements to highlight any cascaded or inhibited emission of one exciton with respect to the others.
 
To verify the nature of each emission line, we perform cross-correlation measurements to highlight any cascaded or inhibited emission of one exciton with respect to the others at 
 $P_{sat}/5$ excitation power (where $P_{sat}$ is the power for exciton saturation). The cross-correlation measurements are done using a transmission spectrometer setup that is aligned to isolate each dot emission line into different paths, which are then sent to SNSPDs connected to a time tagger, as shown in Fig.~\ref{fig1}(b). The results of the cross-correlation measurements are shown in Fig.~\ref{fig2}(c-e). The correlation data between emission lines 1 and 2 shows high bunching around zero positive time delay, indicating the distinct $XX$-$X$  cascade, whereas the $X$ photon is emitted immediately after the $XX$ photon with a very high probability. For correlation (3-1), we get a highly diminished bunching peak (unresolved at the present scale) at a small positive time delay. This indicates that it is the $X^{*}-X$ pair which exhibits negligible cascading. This behaviour is due to the much smaller probability of recapturing a single hole in the dot soon after $\ket{X^{*}}$ recombines to form a $\ket{X}$. Finally, the cross-correlation measurements for (2-3) show a stronger correlation than that of $X^{*}$-$X$, which indicates that it is a $XX-X^{*}$ pair because the nanowire-quantum dots are slightly negatively doped ~\cite{Laferriere_APL2021}, and thus the QD is more readily repopulated with a captured electron than a hole. These cross-correlation measurements enable us to confirm the nature of each emission line in the spectrum, providing critical insights into the behaviour of the excitonic complexes. Line 1 is $X$, line 2 is $XX$ and line 3 is $X^{*}$.

\begin{figure*}[t!]\centering\includegraphics[width=1\linewidth]{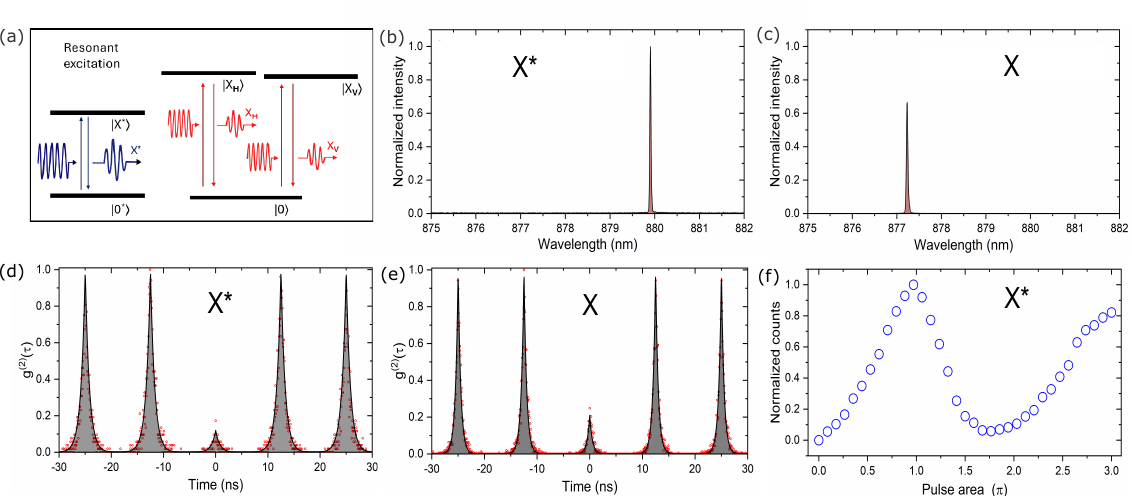}
\caption{ (a) Schematic of energy levels and resonant excitation of charged ($X^{*}$) and neutral exciton ($X$) complexes in a QD. (b and c) PL spectra of a resonantly excited charged exciton and neutral exciton, respectively. (d and e) Second-order correlation ($g^{(2)}(\tau)$)  measurement of charged exciton and neutral exciton, respectively. (f) Normalized count rate as a function of pulse area for resonantly excited charged exciton showing Rabi oscillations up to $3\pi$. }
\label{fig3}
\end{figure*}

After identifying the emission lines, we resonantly excite the neutral and charged exciton (see Fig.~\ref{fig3}(a) for schematic). The pulse duration of the optical excitation is set to 23\,ps using the slicer and then sent to the QD. Fig.~\ref{fig3}(b and c) shows the emission spectra of a resonantly excited charged exciton (at 879.9\,nm) and neutral exciton (at 877.3\,nm) complexes, respectively. To verify the purity of the single photon emission from RF excitation, the fiber-coupled photons are sent to a fiber-based Hanbury Brown Twiss (HBT) setup equipped with two SNSPDs, for second order autocorrelation ($g^{(2)}(\tau)$) measurements. The SNSPDs have efficiencies of 80\% and 66\%  and timing jitters of 18 ps and 11 ps, respectively. The dark counts are less than 10 Hz. A raw and uncorrected $g^{(2)}(0)$ value of 0.078 (see Fig.~\ref{fig3}(d)) was measured for the charged exciton without additional spectral filtering. Similarly, the neutral exciton is analyzed, and a raw $g^{(2)}(0)$ value of 0.21 was obtained (see Fig.~\ref{fig3}(e)). The RMSE for the Trion fitting is $\pm$0.05, while for the exciton,  $\pm$0.02. The higher \( g^{(2)}(0) \) for the neutral exciton arises from less efficient laser rejection, which is not a fundamental limit. The broader linewidth of the neutral exciton requires higher pump power, leading to increased background. Since we focused on the coherent control of the charged exciton, the same pulse width was used for both, resulting in less optimal conditions for the neutral exciton. There may also be contributions from other uncorrected factors like the scattering of the pump light from the tip of the nanowire (the extent of which varies depending on the alignment of the pump light and the NW shape). We measure the RF intensity of the charged exciton as a function of power. The measured data is plotted in Fig.~\ref{fig3}(f), showcasing the Rabi oscillations up to $3\pi$. We observe population inversion of the two-level system at 89\,nW, known as the $\pi$ pulse. As we continue increasing the power, we observe higher-order Rabi oscillations, which correspond to multiple excitation and de-excitation cycles.  

%By varying the excitation power and observing the counts emitted from the charged exciton, the experimental results, showcasing the Rabi oscillations up to $3\pi$, are presented in Fig.~\ref{fig2}(d). In our experiment, we observe a complete population transfer of carriers, known as a $\pi$ pulse, at an excitation power of 89\,nW. As we continue to increase the power, we observe higher-order Rabi oscillations, which correspond to multiple cycles of excitation and de-excitation. 

% NRC RESULTS----------------------------------------------

\begin{figure}[t!]\centering\includegraphics[width=1\linewidth]{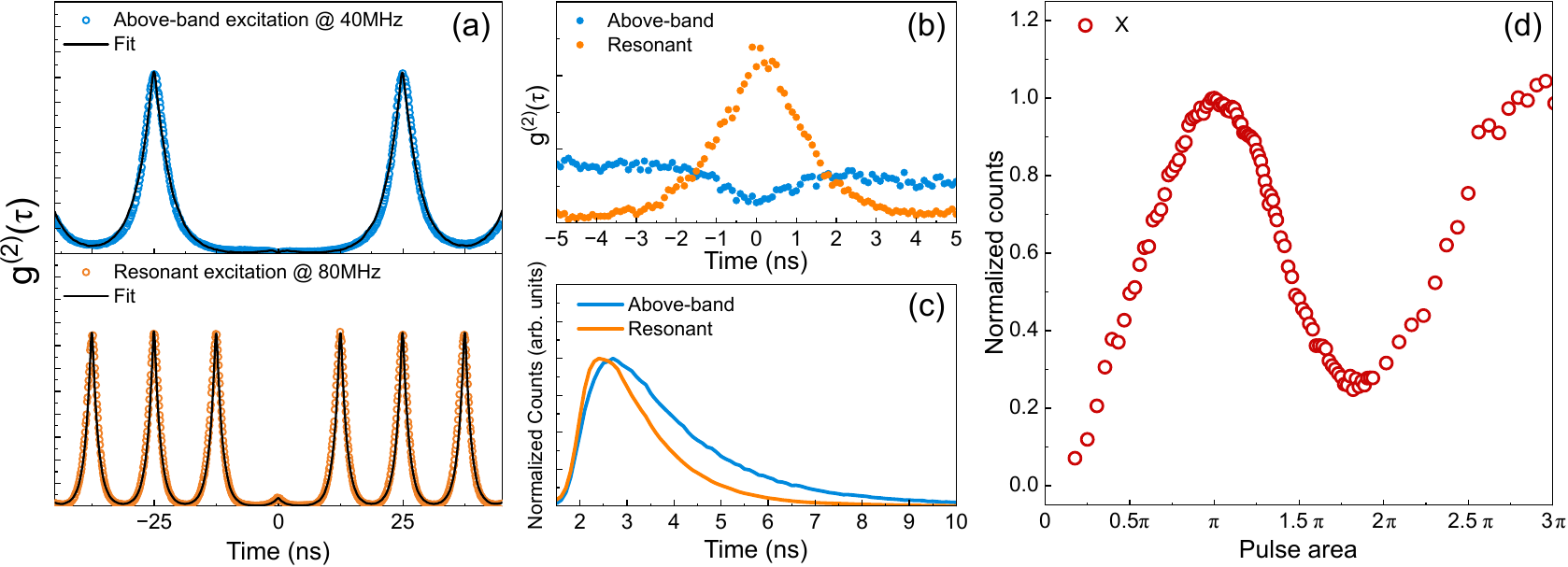}
\caption{(a) Two plots of $g^{(2)} (\tau)$ measurements of neutral exciton emission, fitted with a stochastic model. The results under above-band and resonant excitation are shown in the upper and lower panel correspondingly. (b) A zoom-in plot of $g^{(2)} (\tau)$ around 0 ns time delay, showing the comparison between the two excitation schemes. (c) Lifetime measurements on $X$ under two excitation schemes with the same repetition rate of 80MHz. (d) Rabi oscillations of $X$ under resonant excitation.}
\label{NRC_plots}
\end{figure}

In a comparable setup, we resonantly excite the neutral exciton in another nanowire quantum dot. This quantum dot was grown with slightly adjusted parameters to minimize the exchange splitting in the neutral exciton. A low exchange splitting is preferred for pulsed resonant excitation as it increases the likelihood of resonantly exciting both split levels. The PL intensity of this dot was found to be five times that of the former dot, thereby facilitating time-resolved fluorescence measurements under pulsed above-band and resonant excitation schemes.
\par We measured a $g^{(2)}(0)$ of $0.035 \pm 0.005$ (see Fig.~\ref{NRC_plots}(a)) under pulsed resonant excitation of X as compared to $0.010 \pm 0.005$ when exciting above the band gap at $670$ nm. By looking closely at the zero-delay peak (see Fig.~\ref{NRC_plots}(b)), we gain useful insights regarding the contribution of various factors to the coincidence counts. For the above-band excitation, a dip in the middle with accompanying side plateaus indicates that the main contribution to  $g^{(2)}(0)$ is from the re-excitation of the dot. However, for resonant excitation, there is a peak at zero time-delay where the probability of re-excitation is vanishingly small. This can be explained by the contribution of incoherently scattered laser leaking into the collection path. This laser leakage can be minimized by exact mode-matching of the dark-field microscope with the fundamental mode of the nanowire waveguide. One can also get an upper bound on the residual laser intensity that inadvertently gets mixed with the signal from the height of this peak and post-correct the correlation data. Despite the imperfect laser filtering, we obtain an uncorrected $g^{(2)}(0)<0.07$ for excitation powers up to that corresponding to the $\pi$ pulse (see supplementary). 
Comparing the $g^{(2)}(\tau)$ under above-band and resonant excitation schemes reveals narrower peak widths for resonant excitation, due to a reduced excitation timing jitter associated with the resonant excitation process (see Fig.~\ref{NRC_plots}(a)). From the time-resolved fluorescence counts, we find that there is a sharper rise and a faster decay under resonant excitation ( $T_{1}=1.18\pm0.01 \ ns$) as compared to the above-band excitation (decay time, $\tau=1.83\pm0.01 \ ns$) (see Fig.~\ref{NRC_plots}(c)).  This observation further confirms a reduced excitation timing jitter, i.e., the timing jitter associated with the state preparation process, once the quantum dot is optically pumped for resonant excitation. Since a low-timing jitter is essential for interfering with two identical photons, we expect a considerable improvement in two-photon-interference visibility with resonant excitation. Finally, in Fig.~\ref{NRC_plots}(d), we demonstrate coherent control of X by presenting Rabi oscillations that extend up to $3\pi$.\par 

%To confirm that we are indeed resonantly exciting this dot and rejecting laser, we perform a second order correlation measurement and also record power dependent fluorescence counts. 

The successful demonstration of RF from two distinct nanowire quantum dot samples grown under different conditions highlights the robust compatibility of this excitation technique with nanowire quantum dots. 

\begin{figure*}[t!]\centering\includegraphics[width=1\linewidth]{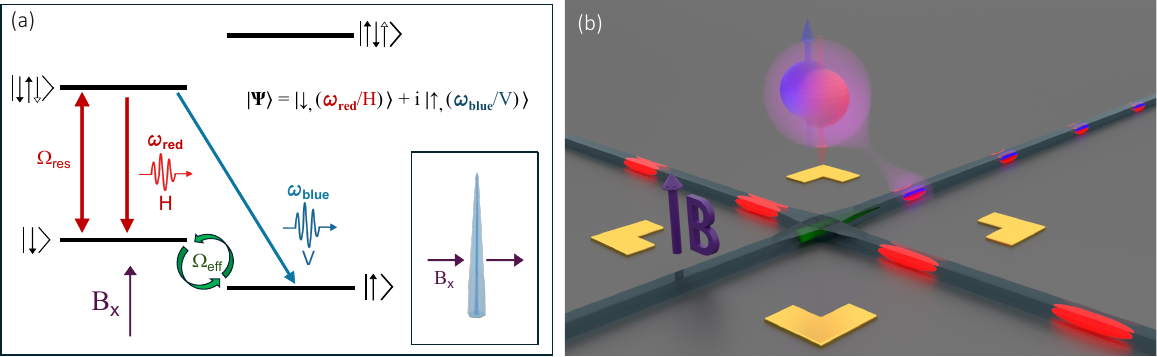}
\caption{NW QD spin photon entanglement. (a) Schematic of spin photon entanglement by means of resonant excitation and application of an external magnetic field. (b) Artistic interpretation of spin photon entanglement in an on-chip waveguide integrated NW QD}
\label{fig5}
\end{figure*}

In conclusion, our work demonstrates the ability to generate single photons on demand and coherently control excitons in nanowire quantum dots. This advancement is pivotal for hybrid quantum photonic systems~\cite{elshaari2020hybrid,elshaari2018strain,elshaari2017chip,gourgues2019controlled}, combining III-V quantum emitters with silicon photonics. These demonstrated results showing coherent control of NW QD emission open up the possibility of exploring on-chip spin photon entanglement and related applications in these systems\cite{Gao2015}. Fig.~\ref{fig5}(a) shows the scheme of spin-photon entanglement in a QD single photon emitter\cite{Gao2012, Gao2015, Schaibley2013}. In presence of an external Voigt magnetic field $B_x$ (B perpendicular to the NW growth axis), a NW QD with a trapped electron charge exhibits Zeeman splitting of its ground state spins and the trion excited states. By applying a resonant excitation pulse of appropriate pulse length, such a system can be initialized to any of the two ground spin states with very high fidelity\cite{Xu2007}. Such high fidelity spin intialization in InAsP/InP NW QDs have already been shown experimentally\cite{Lagoudakis2016}. A QD initialized, to say a $\ket{\uparrow}$ state, when excited resonantly for the $\ket{\uparrow}-\ket{\uparrow\downarrow\Uparrow}$ transition, gets excited and then recombines with a single photon emission to either one of the two ground spin states with equal probabilixtty. The emitted photon is left in a maximally entangled state with the resulting spin state in both the colour (frequency) and polarization (due to transition selection rules) (as illustrated artistically in Fig.~\ref{fig5}(b)). Coherent rotation of the electron spin to any point in the Bloch sphere also can be achieved using off-resonant $\sigma^+$ laser pulses and by utilising the inherent Larmor precession in the applied magnetic field\cite{Berezovsky2008, Press2008}. So, through careful design of the excitation, spin-rotation and readout pulses, advanced spin operations can be performed on such systems. This motivates future exploration of topics like spin-photon teleportation\cite{Gao2013}, spin-spin entanglement\cite{Duan2006} and various quantum repeater architectures based on either quantum memories\cite{Briegel1998} or complex multiphoton entangled states\cite{Azuma2015, Buterakos2017} (all on-chip) on NW QD platforms. The precise positioning and deterministic growth capabilities of NW QDs are also highly advantageous for these causes as it aids the scaling up potential. 

% ---------------------------------------------------------

\begin{acknowledgement}

J.G. acknowledges support from Swedish Research Council (Ref: 2023-06671 and 2023-05288), Vinnova project (Ref: 2024-00466) and the Göran Gustafsson Foundation. E.Y., L.Y. and S.G. acknowledges support from National Sciences and Engineering Research Council of Canada (NSERC). A.W.E acknowledges support from Knut and Alice Wallenberg (KAW) Foundation through the Wallenberg Centre for Quantum Technology (WACQT), Swedish Research Council (VR) Starting Grant (Ref: 2016-03905), and Vinnova quantum kick-start project 2021. V.Z. acknowledges support from the KAW and VR. M.R., S.G. acknowledge support from Mitacs and the National Sciences and Engineering Research Council of Canada (NSERC).

The authors would like to thank Dr. Samuel Gyger for helpful discussions. 

\end{acknowledgement}

\begin{suppinfo}

\end{suppinfo}

\section{Notes}

The authors declare no competing financial interests.

\end{document}